\documentclass[12pt]{article}
\usepackage{latexsym,graphicx}
\newcommand{\be}{\begin{equation}}
\newcommand{\ee}{\end{equation}}
\usepackage[left=2.5cm,top=2.5cm,right=2.5cm,bottom=1.5cm]{geometry}
\usepackage{float}
\usepackage{epstopdf}
\usepackage{cite}
\begin{document}
\begin{center}
\large{\bf{  Traversable Wormholes in $f(R,T)$ Gravity}} \\
\vspace{10mm}
\normalsize{Ambuj Kumar Mishra$^1$, Umesh Kumar Sharma$^2$, Vipin Chandra Dubey$^3$, Anirudh Pradhan$^4$}  \\
\vspace{5mm}
\normalsize{$^{1,2,3,4}$Department of Mathematics, Institute of Applied Sciences and Humanities, GLA University,
Mathura-281 406, Uttar Pradesh, India}\\
\vspace{2mm}
$^1$E-mail: ambuj\_math@rediffmail.com   \\
\vspace{2mm}
$^2$E-mail: sharma.umesh@gla.ac.in  \\
\vspace{2mm}
$^3$E-mail: vipin.dubey@gla.ac.in \\
\vspace{2mm}
$^4$E-mail: pradhan.anirudh@gmail.com\\
\end{center}
\vspace{10mm}
\begin{abstract}
In the present article, models of traversable wormholes within the $f(R, T)$ modified gravity theory are investigated. We have presented some wormhole models, developed from various hypothesis for the substance of their matter, i.e. various relationships with their components of pressure (lateral and radial). The solutions found for the shape functions of the wormholes produced complies with the required metric conditions. The suitability of solution is examined by exploring null, strong and dominant energy conditions. It is surmised that the normal matter in the throat may pursue the energy conditions yet the gravitational field exuding from the adjusted gravity hypothesis support the appearance of the non-standard geometries of wormholes.

\end{abstract}

\smallskip
Keywords: Traversable wormholes, $f(R, T)$ gravity, Energy conditions, Shape functions \\

PACS number: 04.50.kd, 04.20.Gz, 04.20.q

\section{Introduction}
The exact solutions for wormholes (WH) were obtained connecting two distinct regions of asymptotically flat space time in general relativity \cite{ref1,ref2}, also as two distinct anti-de Sitter or asymptotically de-Sitter regions \cite{ref3}. Although, wormholes still have not been observed \cite{ref4} but attempts to achieve has been proposed \cite{ref5,ref6,ref7,ref8,ref9,ref10,ref11,ref12, ref13}. Theoretically wormholes were predicted a long time ago. Wormholes solutions were first proposed by  Einstein and  Rosen \cite{ref14} with event horizon. Many decades later,  Morries and  Thorne have suggested that wormholes can be traversable, which violates the energy conditions, if filled with exotic matter \cite{ref15}.\\ 

In wormholes Physics, the presence of throat is the fundamental feature which satisfy flaring-out conditions \cite{ref16}. Other theories have been also suggested wormhole solutions \cite{ref17,ref18,ref19,ref20,ref21,ref22,ref23,ref24,ref25, ref26} and \cite{ref27} for reviews. NEC for matter is also violated in these works. In fact, the wormhole throat can be threaded by the normal matter in the framework of modified gravity theories and these nonstandard wormhole geometries are supported by the higher order curvature term \cite{ref16}, which can be termed as gravitational fluid. Theoretically, wormholes throats can be formulated in absence of exotic matter \cite{ref28}. Wormholes were constructed by non-minimal coupling in \cite{ref29,ref30}. In modified gravity alone, the fundamental fields sustained the wormhole throats using generic modified gravities \cite{ref31}. Such solutions are also obtained in \cite{ref32,ref33,ref34,ref35,ref36,ref37,ref38}. Rosa et al. \cite{ref16} found wormholes solutions in which the null energy conditions (NEC) was obeyed by the matter not only at throat but everywhere in a hybrid generalized metric-Palatini gravity, with action $f(R, R)$. \\

Indeed, examination of wormhole solutions in various modified theories is a noteworthy and important  point in Theoretical Physics. The General Relativity (GR) could be adjusted in various perspectives. By the introduction of gravitational action of the $f(R, T)$ gravity in Ricci scalar $R$ and the trace of energy momentum tensor $T$, T. Harko and fellow researchers \cite{Harko} streamlined the $f(R)$ theories \cite{refS1,refS2,refS3,refS4,refS5,refS6} by replacing the function $f(R)$ with a random function $f(R, T)$. This $f(R, T)$ theory tested  in Astrophysics of compact objects \cite{refN1,refN2,refN3,refN4}, Thermodynamics \cite{refN5,refN6} and cosmology \cite{refN7,refN8,refN9,refN10,refN11,refN12,refN13,refN14,refN15,refN16,refN17,refN18}. \\

The work has been done on the cases of $WH$ geometry in $f(R, T)$ gravity , where the redshift function $\phi$ is not dependent on either time or spatial coordinate \cite{refn19,refn20}. The spherically symmetric and static  wormholes are considered in $f(R,T) = f(R) + \Lambda T$ gravity with different fluids  \cite{refn21}. The geometry (non-commutative) of wormholes with respect to Lorentzian and Gaussian   distributions of string theory is proposed  as well as found the numerical and exact  solutions in modified $f(R,T)$ gravity \cite{refn22}.  An analytical approach is used to get the wormhole solutions in $f(R,T)$ gravity  \cite{refn23}. An exponential shape function is defined for wormholes and a comparison is made between $f(R,T)$ and  $f(R)$ gravity for the rationality of energy conditions \cite{refn24}.  Wormholes are studied  for two types of varying Chaplygin gas and found the violation of dominated and null energy conditions in $f(R,T)$ gravity \cite{refn25}.  In \cite{refn26}, an exponential $f(R,T)$ gravity model has been considered and obtained solutions for wormhole. The charged wormhole solutions are investigated in $f(R,T)$ gravity  satisfying the energy conditions \cite{refn27}. Using Noether symmetry approach, wormhole solutions  were investigated for non dust and dust distributions  in $f(R,T)$ gravity \cite{refn28}. Considering the different types of energy density,  traversable wormhole solutions investigated in $f(R,T)$ gravity \cite{refn29}. Wormhole solutions have been studied by various researchers in modified gravity theories in various set-ups \cite{refa1,refa2,refa3,refa4,refa5}.\\
 
 The fact that wormholes matter content is signified by an anisotropic fluid and $T$- dependence of the $f(R, T)$ theory, which may be related to imperfect fluid forms the strong motivation behind working with the wormholes in $f(R, T)$ gravity \cite{refn23,refa6}. Recently,  traversable wormhole are investigated for two particular types of shape functions and obtained wormhole structures filled with phantom fluid \cite{ref59}. Further, traversable wormholes  investigated the  in $f(R,T)$ gravity by taking a new form of $f(R,T) = R +2αlnT$ function to minimize the appearance of exotic  matter near the throat of the wormhole \cite{refa7}.\\
 
  Hence, in this paper, we have investigated the traversable wormholes  by assuming functional  form of  $f(R,T) = R +\Lambda T$ of  $f(R,T)$ gravity, originally proposed by the authors \cite{Harko} considering the shape function given in \cite{refa6}, which has not been studied already. \\
  
  The paper is organized as follows: in Section II, a brief review is  presented for the $f(R,T)$ theory. In Section III, the wormhole metric and its conditions are described.  In Section IV, we presented the field equation solution for the wormhole metric in the $f(R,T)$ gravity. In Section V, we presented models of the wormhole. Our results are discussed in Section VI.
    
\section{The basic formalism of $f(R, T)$ gravity} 
The theory of  $f(R, T)$ gravity is given by the following action \cite{Harko}

\begin{equation}\label{eq1}
S = \frac{1}{16 \pi} \int d^4 x  f(R, T) \sqrt{-g} + \int \mathcal{L}_m d^4 x \sqrt{-g},  
\end{equation}
where $R$ is Ricci scalar and $T$ is trace of the energy momentum tensor (EMT) in an  arbitrary function $f(R, T)$, $g$ is the metric determinant and matter Lagrangian density is $\mathcal{L}_m $. These are connected to the EMT \cite{ref39}.

\begin{equation}\label{eq2}
T_{ij} = -\frac{2}{\sqrt{-g}} \left[\frac{\partial(\sqrt{-g}\mathcal{L}_m)}{\partial g^{ij}}- \frac{\partial}{\partial x^k} \frac{\partial(\sqrt{-g}\mathcal{L}_m)}{\partial  
(\frac{\partial g^{ij}}{\partial x^k})}\right].
\end{equation}
Moreover, units is taken such that $c = 1 = G$. 

Following as \cite{ref40,Harko}, we consider $\mathcal{L}_m$ relies only on $g_{ij}$ (metric component) and not depends on its differential coefficients, such that we can get
\begin{equation}\label{eq3}
T_{ij} = - \frac{\partial \mathcal{L}_m}{\partial g^{ij}} + g_{ij}\mathcal{L}_m. 
\end{equation}

In the Eq. (\ref{eq1}),  the action $S$ is varying with  $g_{ij}$ which provides the $f(R, T)$ field equations\cite{Harko} as

$f_R (R, T)\left(R_{ij}- \frac{1}{3}R g_{ij}+ \frac{1}{6} f(R, T) g_{ij}\right)$\\

\qquad\qquad $= 8\pi G \left(T_{ij}- \frac{1}{2} T g_{ij}\right) - f_T (R, T)\left(T_{ij}-\frac{1}{3} T g_{ij}\right)$\\
\begin{equation}\label{eq4}
-f_T (R, T)\left(\theta_{ij}-\frac{1}{3}\theta g_{ij}\right)+ \bigtriangledown_i \bigtriangledown_j f_R(R, T)
\end{equation}
with $f_R(R,T) = \frac{\partial f(R, T)}{\partial R}, f_T(R, T)= \frac{\partial f(R, T)}{\partial T}$ and 
\begin{equation}\label{eq5}
\theta_{ij} = g^{ij}\frac{\partial T_{ij}}{\partial g^{ij}}.
\end{equation}
Such theories of gravity $f(R, T)$ (as well as $f(R, \mathcal{L}_ m)$) can explain in their formalism a non-minimal matter-geometry coupling.  It suggests that research particles in the field (gravitational) are not going to obey geodesic lines in such theories. The  geometry and matter  coupling prompts an additional force one behaves perpendicular to the four-velocity on the particles.\\ 

Interestingly, an additional force relies on the Lagrangian structure of the matter \cite{ref42}. It has been found in \cite{ref43},  by taking the total pressure of $\mathcal{L}_m = p$ with $p$, the additional force disappear. However, more normal structures of $\mathcal{L}_ m$ such as $\mathcal{L}_m = -\rho$ with $\rho$ represents the density of energy are more common apparently and  do not give the vanishing of the additional force. 

 Assuming the Lagrangian matter in this paper as $\mathcal{L}_m = -\rho$. It is therefore possible to write an equation (\ref{eq5}) as  
\begin{equation}\label{eq6}
\theta_{ij} = -2T_{ij}-\rho g_{ij}.
\end{equation}
Let us suppose the function $f(R, T) = 2f(T) + R$, where $f(T)$ is a random function of $T$.  The field equations (\ref{eq4}) and (\ref{eq5} ) of $f(R, T)$ gravity takes the form
\begin{equation}\label{eq7}
R_{ij}-\frac{1}{2}R g_{ij} = 8\pi T_{ij} + 2f^t(T)T_{ij}+[2\rho f^t(T)+f(T)]g_{ij}, 
\end{equation} 
where $R_{ij}$ is Ricci tensor and $f^t(T)= \frac{df(T)}{dT}$. By taking $f(T)=\lambda T$, with a constant $\lambda$, we can rewrite the above as
\begin{equation}\label{eq8}
G_{ij}= \left(8\pi + 2 \lambda \right)T_{ij} +\lambda \left(2\rho +T\right)g_{ij},
\end{equation}
where $G_{ij}$ is usual Einstein tensor. Such a $f(R, T)$ functional type assumption was originally proposed in \cite{Harko}. 
\section{Wormhole Metric and Its Conditions}
In Schwarzchild coordinate $\left(t, r, \theta, \phi\right)$, the spherical symmetric wormhole metric is \cite{ref44, ref55} 
\begin{equation}\label{eq9}
ds^2= - U(r)dt^2+ \frac{dr^2}{V} + r^2 d\Omega^2,
\end{equation}
where $\Omega^2 = d\theta^2 + sin^2\theta d\phi^2$ and $V=1-\frac{b(r)}{r}$. The $U(r)$ that denotes redshift function and the $b(r)$ that denotes shape function given in Eq. (\ref{eq9})  will follow the conditions mentioned below \cite{ref44, ref55}
\begin{enumerate}
	\item The coordinate  $r$  named as radial co-ordinate,  ranges $r_0 \leq r\leq \infty$, where $r_0$ refers to the radius of the throat.
	\item  The shape function ($b(r)$) fulfills the condition at the throat
	\begin{equation}\label{eq10}
	b(r_0)= r_0,
	\end{equation}
	and for the out of throat i.e. for $r>r_0$
	\begin{equation}\label{eq11}
	0 < 1-\frac{b(r)}{r}.
	\end{equation}
	\item The flaring out condition must be satisfied by shape function $b(r)$  at the throat, i.e.
	\begin{equation}\label{eq12}
	b'(r_0) <1,
	\end{equation}
	where superscript ($ ' $) represent derivative with respect to $r$
	\item The limit required for asymptotically flatness of the space-time geometry
	\begin{equation}\label{eq13}
	\frac{b(r)}{r}\rightarrow 0  \qquad as \qquad |r|\rightarrow \infty.
	\end{equation}
	\item At the throat $r_0$, redshift function $U(r)$ must be non-vanishing and finite. \\
\end{enumerate}
To achieve the anti-de Sitter and de Sitter asymptotic behavior, we can take $U(r)$= constant. Since the re-scaled time coordinate can absorb a constant redshift, we take $U(r)=1$, such as \cite{ref57, ref58}.

\begin{figure}
	\centering 
	(a)\includegraphics[width=12cm, height=8cm, angle=0]{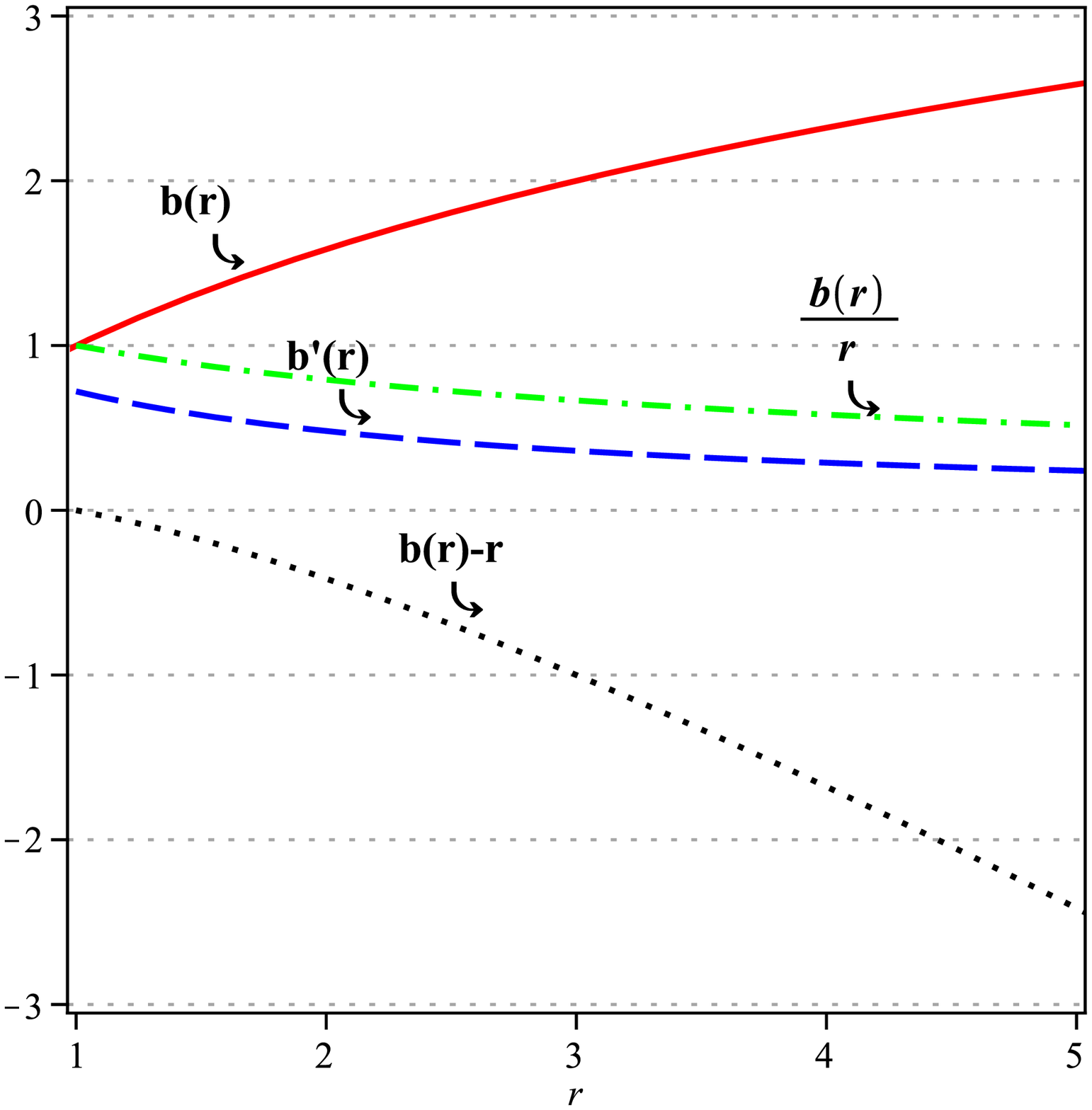}
	\caption {Nature of $b(r)$, throat condition $b(r)<r$, flaring out condition $b'(r)<1$ and asymptotically flatness $\lim\limits_{r\rightarrow \infty} \frac{b(r)}{r}=0$ for $r_0 = 1$.}
\end{figure} 

\begin{figure}
	(a)\includegraphics[width=8cm, height=8cm, angle=0]{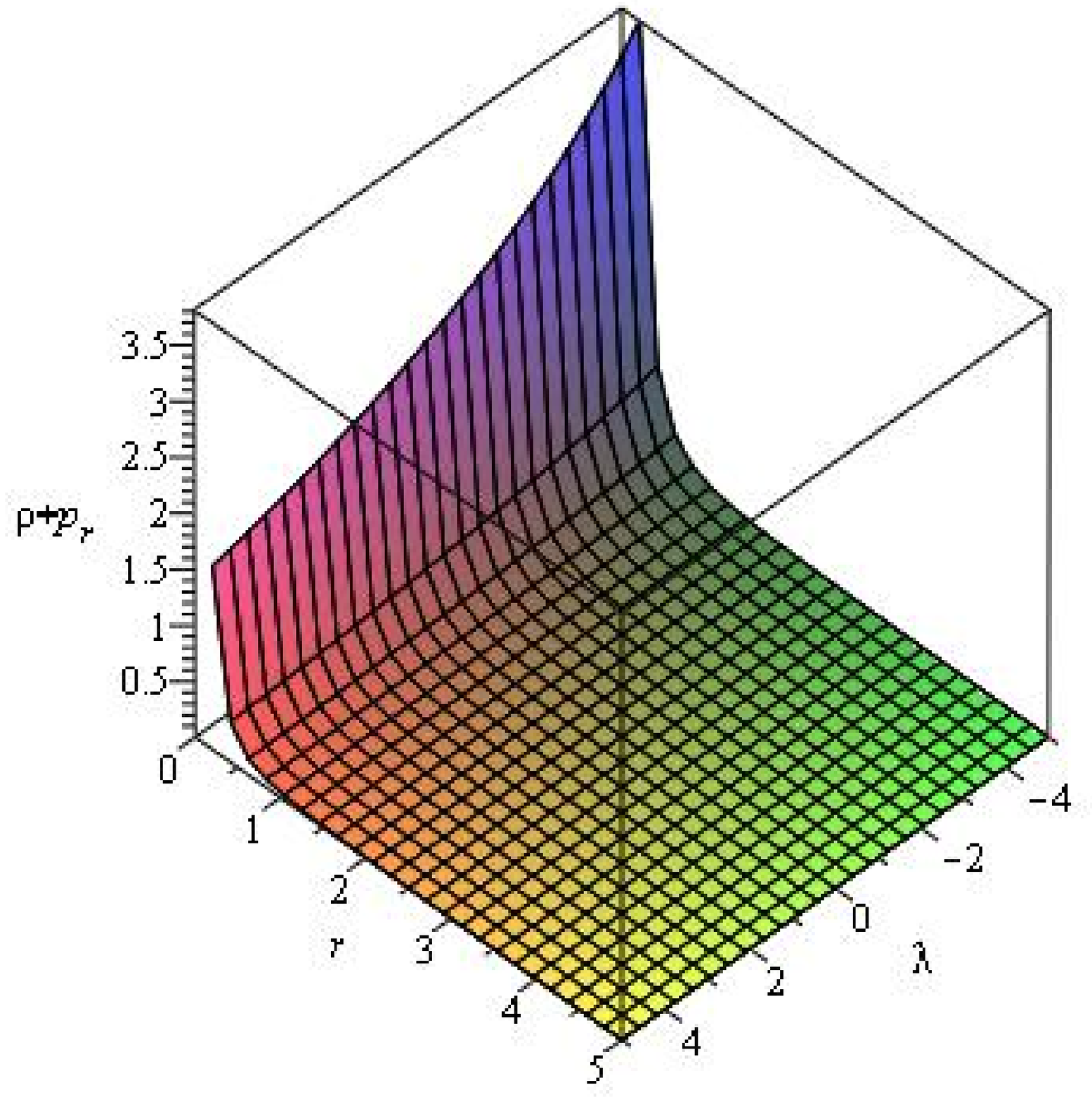}
	(b)\includegraphics[width=8cm, height=8cm, angle=0]{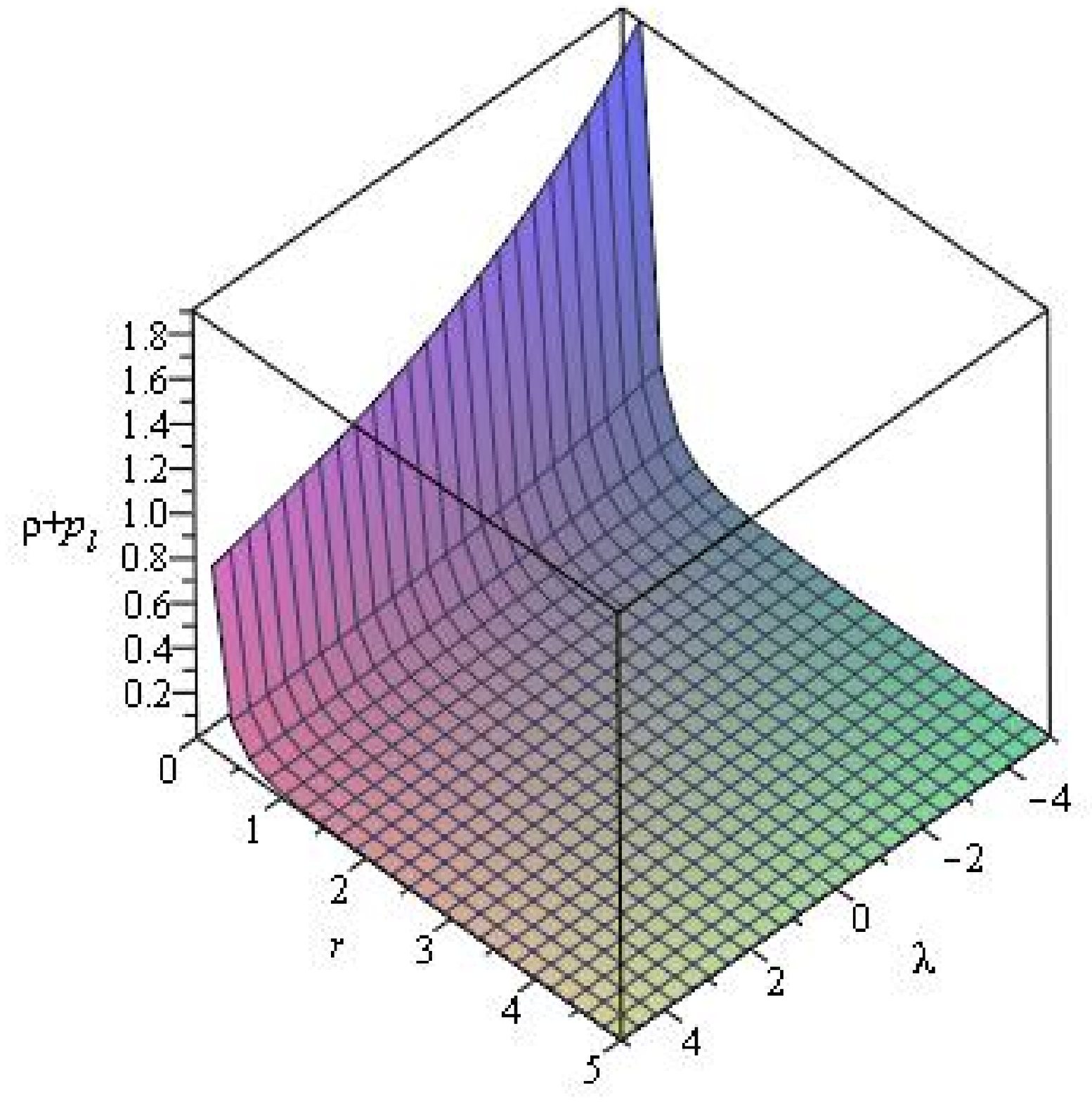}
	\caption {(a) NEC, $\rho + p_r$ , (b) NEC, $\rho + p_l$ .}
\end{figure}
\begin{figure}
	(a)\includegraphics[width=8cm, height=8cm, angle=0]{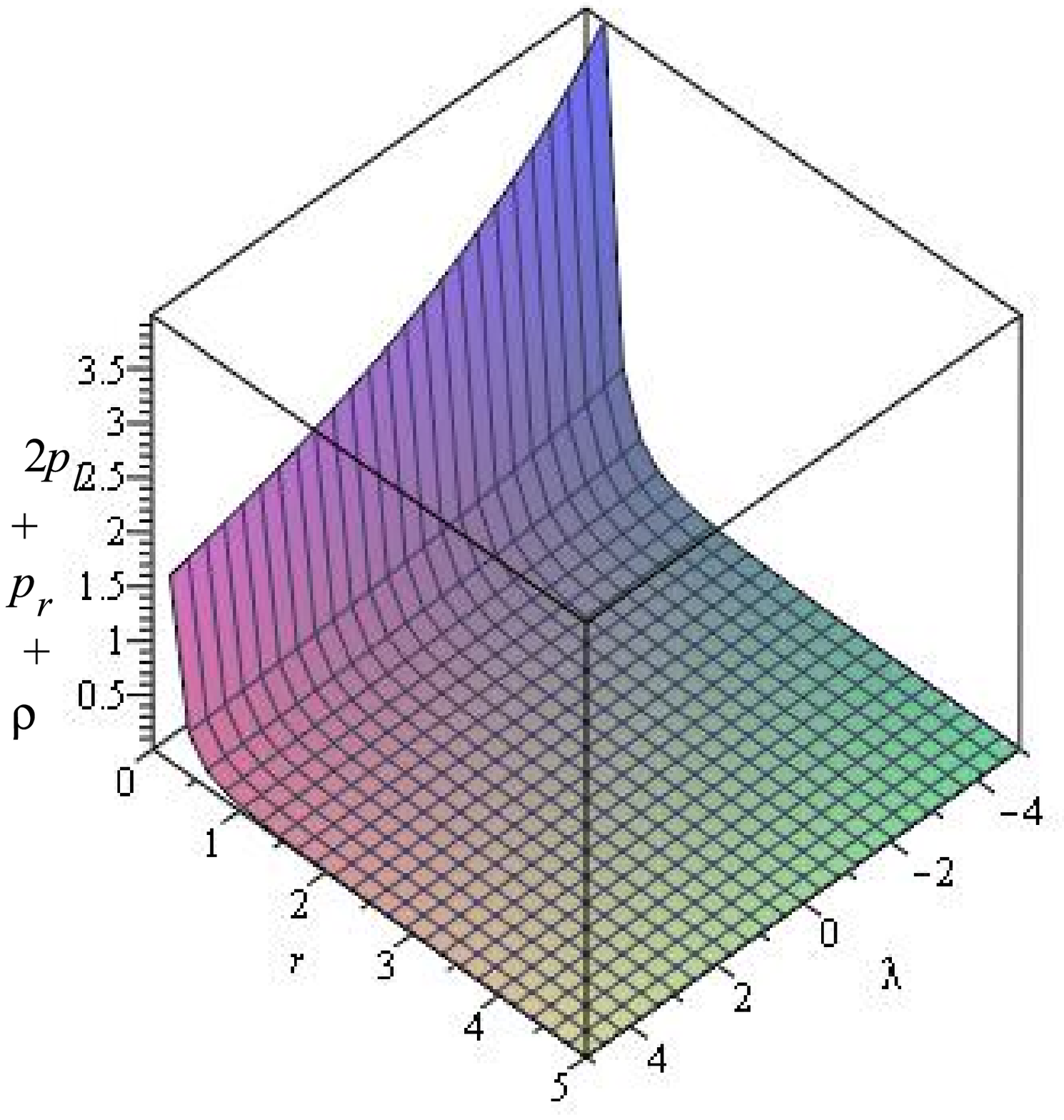}
	(b)\includegraphics[width=8cm, height=8cm, angle=0]{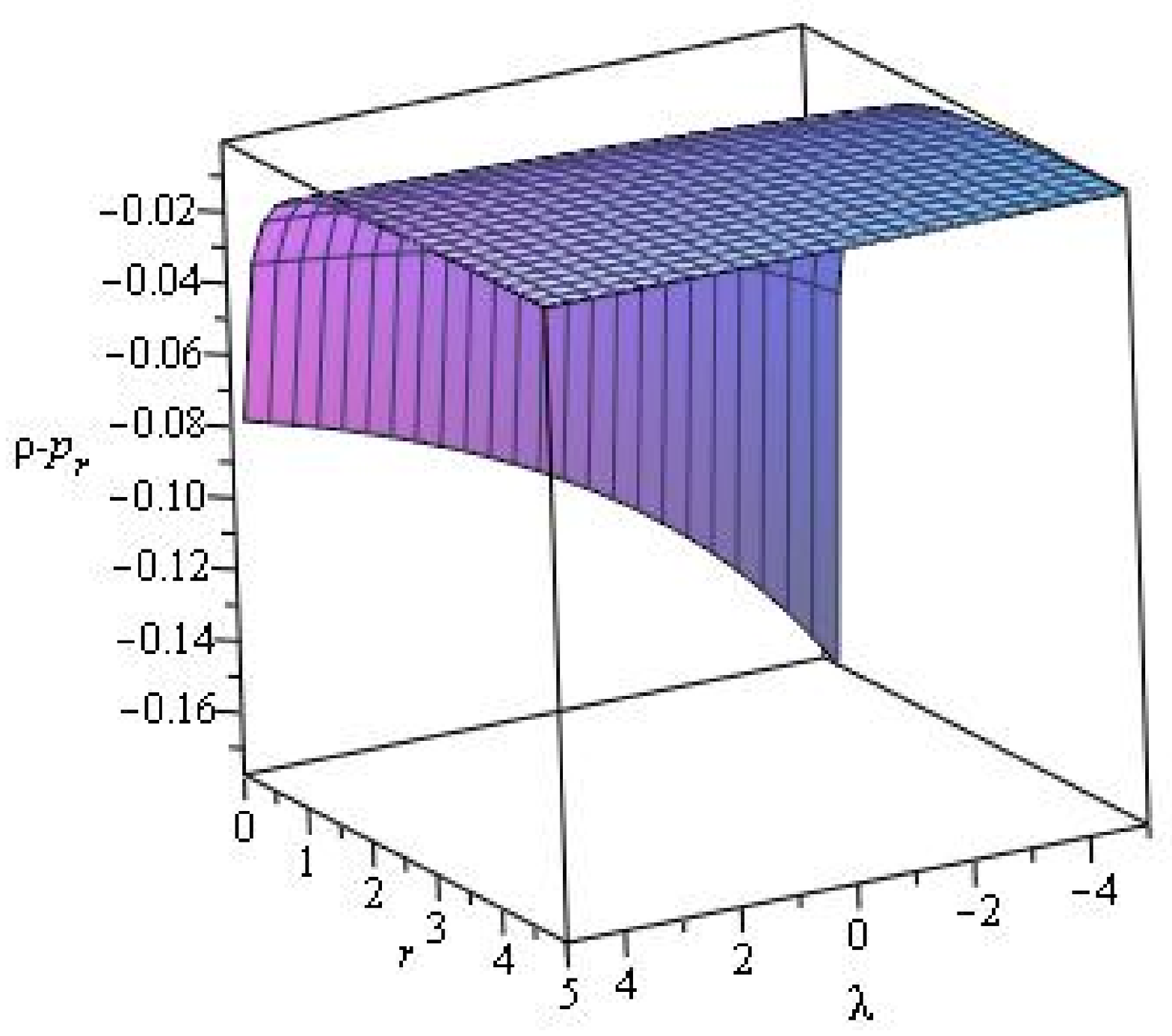}
	\caption {(a) SEC, $\rho + p_r + 2p_l$, (b) DEC, $\rho - p_r$.}
\end{figure}
\begin{figure}
	(a)\includegraphics[width=8cm, height=8cm, angle=0]{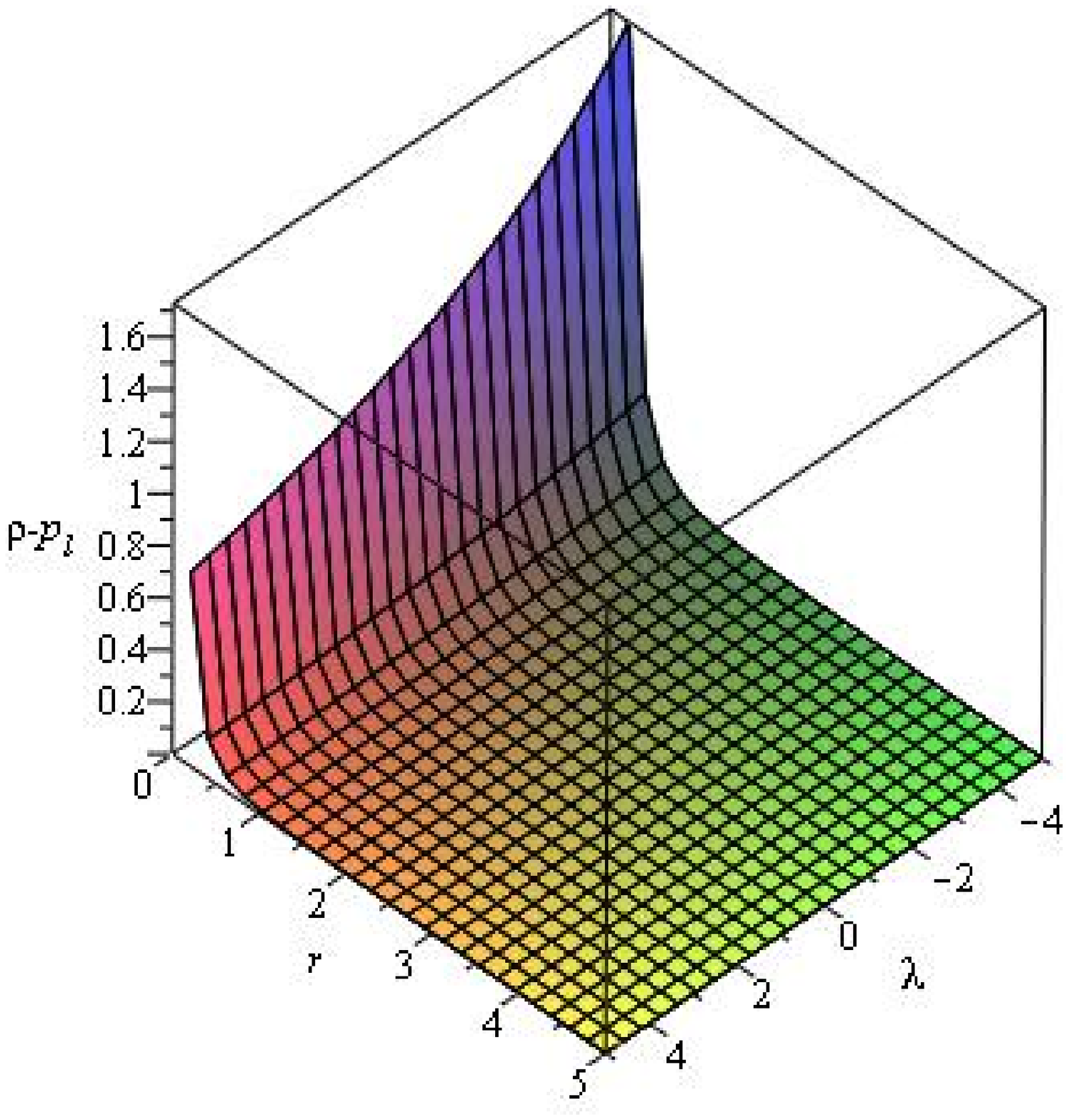}
	(b)\includegraphics[width=8cm, height=8cm, angle=0]{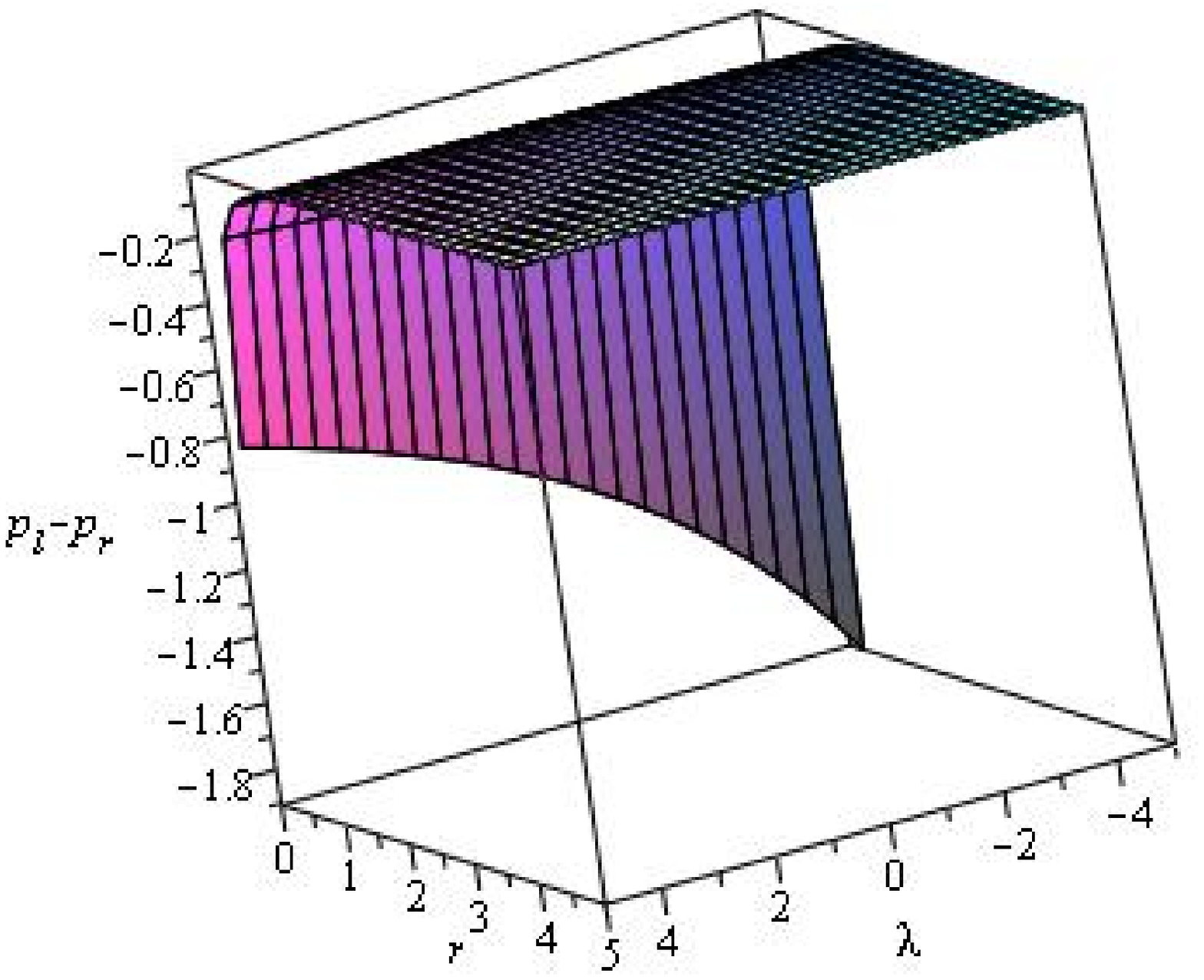}
	\caption {(a) DEC, $\rho - p_l$, (b) DEC, $p_l - p_r$.}
\end{figure}

\section{Solutions of Field Equations for Wormholes in $f(R, T)$ Gravity.}

Considering that the description of matter is described by an anisotropic fluid of the form 
\begin{equation}\label{eq14}
T^i _j = diag \left\lbrace-\rho, p_r, p_l, p_l\right\rbrace.
\end{equation}
Where $\rho = \rho(r), p_r = p_r(r), p_l = p_l(r)$ are energy density, radial pressure and transverse (measured orthogonal to radial directions) pressure, respectively and the trace $T$ of the Eq. (\ref{eq14}) can be obtained as $T=-\rho + p_r + 2p_l$.\\

The  field equation components from Eq.  (\ref{eq8}) for the metric given by Eq.  (\ref{eq9}) with Eq. (\ref{eq14}) are obtained as :
\begin{equation}\label{eq15}
\frac{b'}{r^2} = \left(8\pi + \lambda\right)\rho - \lambda \left(p_r+ 2p_l\right),
\end{equation}
\begin{equation}\label{eq16}
-\frac{b}{r^3} =\lambda \rho + \left(8\pi +3\lambda\right)p_r + 2\lambda  p_l,
\end{equation}
\begin{equation}\label{eq17}
\frac{b-b'r}{2r^3} = \lambda \rho + \lambda p_r +\left(8\pi + 4\lambda\right)p_r,
\end{equation}
the  field equations given by Eqs. (\ref{eq15}), (\ref{eq16}) and(\ref{eq17}), yields the following solutions
\begin{equation}\label{eq18}
\rho = \frac{b'}{r^2 (8\pi + 2\lambda)},
\end{equation}
\begin{equation}\label{eq19}
p_r = -\frac{b}{r^3 (8\pi + 2\lambda)},
\end{equation}  
\begin{equation}\label{eq20}
p_l = \frac{b-b'r}{2r^3 (8\pi + 2\lambda)}.
\end{equation}


\section{Models of Wormholes}
 We shall investigate models of wormholes for two different shape functions in this section.
\subsection{Shape Function $b(r)= \frac{r_0 \log(r+1)}{\log(r_0 +1)}$}
For our first model we take radial pressure $p_r$ and energy density $\rho$  related as 
\begin{equation}\label{eq21}
	p_r = \omega \rho,
\end{equation}
where equation of state parameter $\omega$ in terms of $p_r$ is called radial state parameter.  
The $f(R, T)$  gravity model is considered with the shape function \cite{ref59} 
\begin{equation}\label{eq22}
	b(r)= \frac{r_0 \log(r+1)}{\log(r_0 +1)}.
\end{equation}

For this shape function, wormholes solutions are obtained. The energy density, radial pressure, tangential pressure can be obtained by using field equations (\ref{eq18})- (\ref{eq20}) as
\begin{equation}\label{eq23}
	\rho ={\frac {r_0}{ \left( 1+r \right) \ln  \left(1 +r_0
			\right) {r}^{2} \left( 8\,\pi +2\,\lambda \right) }},
\end{equation}  
\begin{equation}\label{eq24}
	p_r = {\frac {r_0 \ln  \left( 1+r \right) }{\ln  \left(r_0 +1
			\right) {r}^{3} \left( 8\,\pi +2\,\lambda \right) }},
\end{equation}

\begin{equation}\label{eq25}
	p_l = 1/2  \left( {\frac {r_0 \ln  \left( 1+r \right) }{\ln  \left( r_0 +1 \right)}}-{\frac {r_0  r}{ \left( 1+r \right) \ln 
			\left( 1 +r_0 \right) }} \right) {r}^{-3} \left( 8 \pi +2 \lambda \right) ^{-1},
\end{equation}
\begin{equation}\label{eq26}
	\rho + p_r = \frac{1}{2}{\frac {r_0 \, \left( r+r\ln  \left( 1+r \right) +\ln \left( 1+r \right)  \right) }{ \left( 1+r \right) \ln  \left( r_0 +1 \right) {r}^{3} \left( 4\,\pi +\lambda \right) }},
\end{equation}
\begin{equation}\label{eq27}
	\rho + p_l = \frac{1}{4}{\frac {r_0\, \left( r+ r \ln  \left( 1+r \right) +\ln \left( r+1 \right)  \right) }{ \left( 1+r \right) \ln  \left( r_0 +1 \right) {r}^{3} \left( 4\,\pi +\lambda \right) }},
\end{equation}
\begin{equation}\label{eq28}
	\rho + p_r + 2p_l = {\frac {r_0 \,\ln  \left(1+r \right) }{\ln  \left(r_0 +1 \right) {r}^{3} \left( 4\,\pi +\lambda \right) }},
\end{equation}
\begin{equation}\label{eq29}
	\rho - |p_r|= {\frac {r_0 }{ \left( 1+r \right) \ln  \left(r_0 +1
			\right) {r}^{2} \left( 8\,\pi +2\,\lambda \right) }}- \left| {\frac {r_0 \,\ln  \left( 1+r \right) }{\ln  \left(r_0 +1 \right) {
				r}^{3} \left( 8\,\pi +2\,\lambda \right) }} \right|,
\end{equation}
\begin{equation} \label{eq30}
	\rho - |p_l| ={\frac {r_0 }{ \left( 1+r \right) \ln  \left(r_0 +1
			\right) {r}^{2} \left( 8\,\pi +2\,\lambda \right) }}-\frac{1}{2} \left| \left( {\frac {r_0 \,\ln  \left( 1+r \right) }{\ln  \left( 
			r_0 +1 \right) }}-{\frac {rr_0 \,}{ \left( 1+r \right) \ln 		\left( r_0 +1 \right) }} \right) {r}^{-3} \left( 8\,\pi +2\,
	\lambda \right) ^{-1} \right|, 
\end{equation}
\begin{equation}
	p_l - p_r = -\frac{1}{4}{\frac {r_0 \, \left( r+\ln  \left(1+r \right) r+\ln \left( 1+r \right)  \right) }{ \left( 1+r \right) \ln  \left( r_0 +1 \right) {r}^{3} \left( 4\,\pi +\lambda \right) }}.
\end{equation} 

We can see directly from Fig. 1, that all the conditions i.e. fundamental needs of shape functions, such as throat condition, flaring out condition and typical asymptotically flatness condition are satisfied for WH geometry. It can be observed from Eqs. (\ref{eq26}, \ref{eq27})  and Figs. 2(a) and 2(b) the validation region of null energy condition (NEC) i.e. $\rho + p_r \geq 0$ and  $\rho + p_l \geq 0$. This is the weakest restriction and just represents the attractive nature of gravity.\\

The strong energy condition (SEC) derives from the attractive nature of gravity and its shape arises directly from the analysis of a spherically symmetrical metric in the GR system. From (\ref{eq28}) the SEC is plotted in Fig. 3(a). The SEC $\rho + p_r +2p_l$ is found to  be positive, decreasing and tending towards zero with the increment of $r$.\\

The dominant energy condition (DEC) limited the rate of the transfer of energy to the speed of light. DECs are obtained in Eqs. (\ref{eq29}, \ref{eq30}). DEC is plotted in Figs 3(b) and 4(a). one can observe that violation of $\rho \geq |p_r| $ for present model. We also observe from Fig. 4(b), that anisotropy parameter is negative with change in $r$, which signifies  that the geometry is of attractive nature.   

  
  \subsection{Shape Function $b(r)= r_0 \left(\frac{r}{r_0}\right)^\gamma , \,\,\, 0<\gamma<1.$}
In this model, the wormholes solutions are obtained by using the shape function  $b(r)= r_0 \left(\frac{r}{r_0}\right)^\gamma , \,\,\, 0<\gamma<1$ (see in \cite{ref59}). In this case energy density $\rho$, radial pressure $p_r$, tangential pressure $p_l$  are obtained by using field equations (\ref{eq18})-(\ref{eq20})
\begin{equation}\label{eq32}
	\rho =r_0 \left( {\frac {r}{r_0}} \right) ^{\gamma}\gamma{r}^{
		-3} \left( 8\,\pi +2\,\lambda \right) ^{-1}
\end{equation}  
\begin{equation}\label{eq33}
	p_r = r_0 \, \left( {\frac {r}{r_0 }} \right) ^{\gamma}{r}^{-3}
	\left( 8\,\pi +2\,\lambda \right) ^{-1}
\end{equation}  
\begin{equation}\label{eq34}
	p_l = 1/2\, \left(r_0 \, \left( {\frac {r}{r_0 }} \right) ^{
		\gamma}-r_0 \, \left( {\frac {r}{r_0 }} \right) ^{\gamma}
	\gamma \right) {r}^{-3} \left( 8\,\pi +2\,\lambda \right) ^{-1}
\end{equation}  
\begin{equation} \label{eq35}
	\rho + p_r = r_0 \, \left( {\frac {r}{r_0 }} \right) ^{\gamma}\gamma{r}^{-3} \left( 8\,\pi +2\,\lambda \right) ^{-1}+r_0 \, \left( {\frac {r}{r_0 }} \right) ^{\gamma}{r}^{-3} \left( 8\,\pi +2\,\lambda   \right) ^{-1}
\end{equation}

\begin{equation}\label{eq36}
	\rho- p_l = r_0 \, \left( {\frac {r}{r_0 }} \right) ^{\gamma}\gamma{r}^{-3} \left( 8\,\pi +2\,\lambda \right) ^{-1}+1/2\, \left(r_0 \,\left( {\frac {r}{r_0 }} \right) ^{\gamma}-r_0 \, \left( { \frac {r}{r_0 }} \right) ^{\gamma}\gamma \right) {r}^{-3} \left( 8\,\pi +2\,\lambda \right) ^{-1}
\end{equation}
$$\rho +p_r + 2p_l =r_0 \, \left( {\frac {r}{r_0 }} \right) ^{\gamma}\gamma{r}^{-3} \left( 8\,\pi +2\,\lambda \right) ^{-1}+r_0 \, \left( {\frac 	{r}{r_0 }} \right) ^{\gamma}{r}^{-3} \left( 8\,\pi +2\,\lambda \right) ^{-1}$$
\begin{equation}\label{eq37}
	+ \left( r_0 \, \left( {\frac {r}{r_0 }}
	\right) ^{\gamma}-r_0 \, \left( {\frac {r}{r_0 }} \right) ^
	{\gamma}\gamma \right) {r}^{-3} \left( 8\,\pi +2\,\lambda \right) ^{-1}
\end{equation}
\begin{equation}\label{eq38}
	\rho - |p_r| =r_0 \, \left( {\frac {r}{r_0 }} \right) ^{\gamma}\gamma{r}^{-3} \left( 8\,\pi +2\,\lambda \right) ^{-1}- \left| r_0 \, \left( {\frac {r}{r_0 }} \right) ^{\gamma}{r}^{-3} \left( 8\, \pi +2\,\lambda \right) ^{-1} \right| 
\end{equation}
\begin{equation}\label{eq39}
	\rho - |p_l| = r_0 \, \left( {\frac {r}{r_0 }} \right) ^{\gamma}\gamma{r}^{-3} \left( 8\,\pi +2\,\lambda \right) ^{-1}-1/2\, \left|  \left( r_0 \, \left( {\frac {r}{r_0 }} \right) ^{\gamma}-r_0 \, \left( {\frac {r}{r_0 }} \right) ^{\gamma}\gamma \right) {r}^{-3} \left( 8\,\pi +2\,\lambda \right) ^{-1} \right| 
\end{equation}
\begin{equation}\label{eq40}
	p_l - p_r = 1/2\, \left( r_0 \, \left( {\frac {r}{r_0 }} \right) ^{
		\gamma}-r_0 \, \left( {\frac {r}{r_0 }} \right) ^{\gamma}
	\gamma \right) {r}^{-3} \left( 8\,\pi +2\,\lambda \right) ^{-1}-
	r_0 \, \left( {\frac {r}{r_0 }} \right) ^{\gamma}{r}^{-3}
	\left( 8\,\pi +2\,\lambda \right) ^{-1}
\end{equation}

We can also see from Fig. 5 that all the conditions mentioned earlier for WH geometry are satisfied for second model. Further, in Figs 6(a) and 6(b) we plot the validity region of NEC $(\rho >o)$. The validity of SEC can also be seen in Fig. 7(a). One can observe from Fig. 7(b) that violation of DEC ($\rho > |p_r|$). $\rho-p_{l}$ is positive as $r$ increases shown in Fig. 8(a). It can be seen from Fig. 8(b), that anisotropy parameter is negative with change in $r$, which signifies  that the geometry is of attractive nature.   
 
 \begin{figure}
	\centering
	(a)\includegraphics[width=12cm, height=8cm, angle=0]{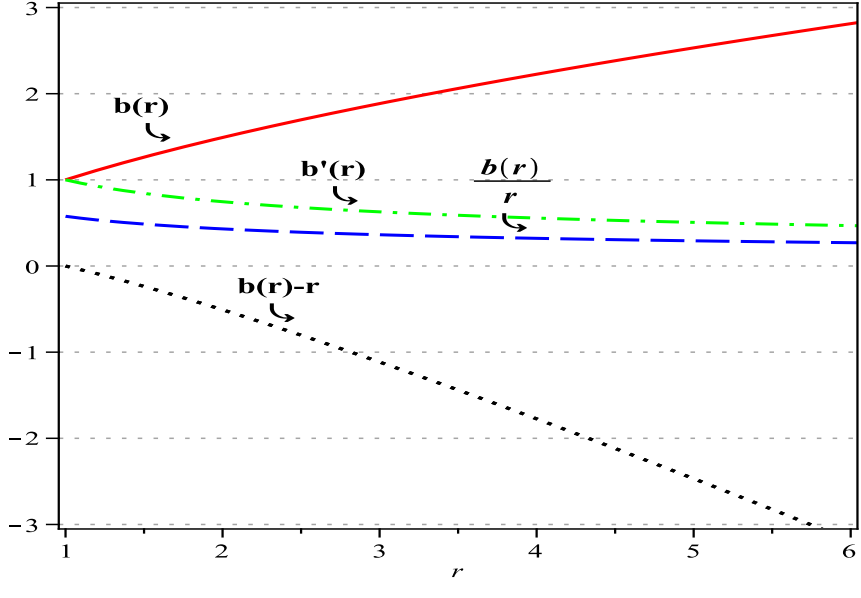}
	\caption {Nature of $b(r)$, throat condition $b(r)<r$, flaring out condition $b'(r)<1$ and asymptotically flatness $\lim\limits_{r\rightarrow \infty} \frac{b(r)}{r}=0$ for $r_0 = 1$.}
\end{figure}

\begin{figure}
	(a)\includegraphics[width=8cm, height=8cm, angle=0]{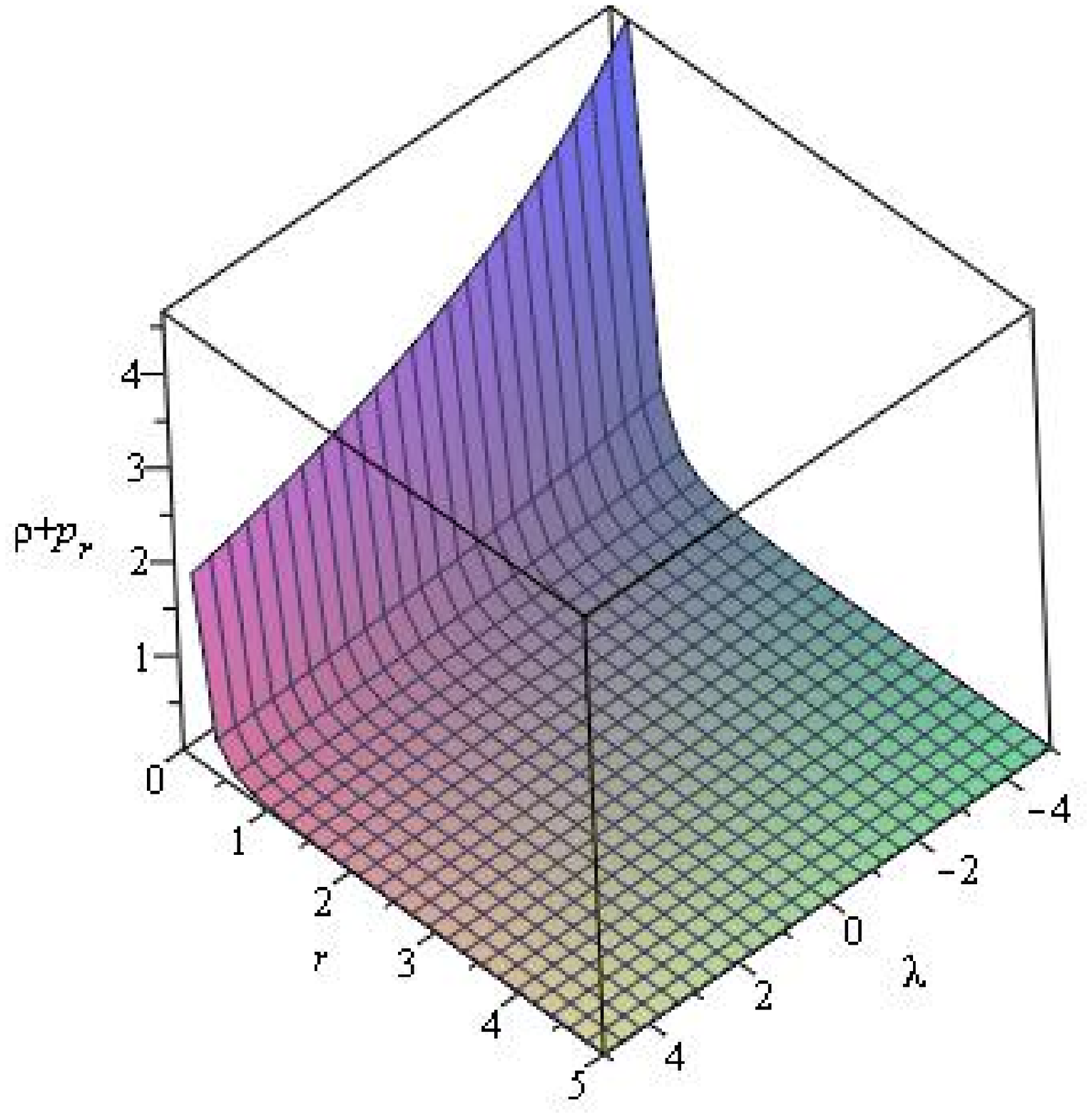}
	(b)\includegraphics[width=8cm, height=8cm, angle=0]{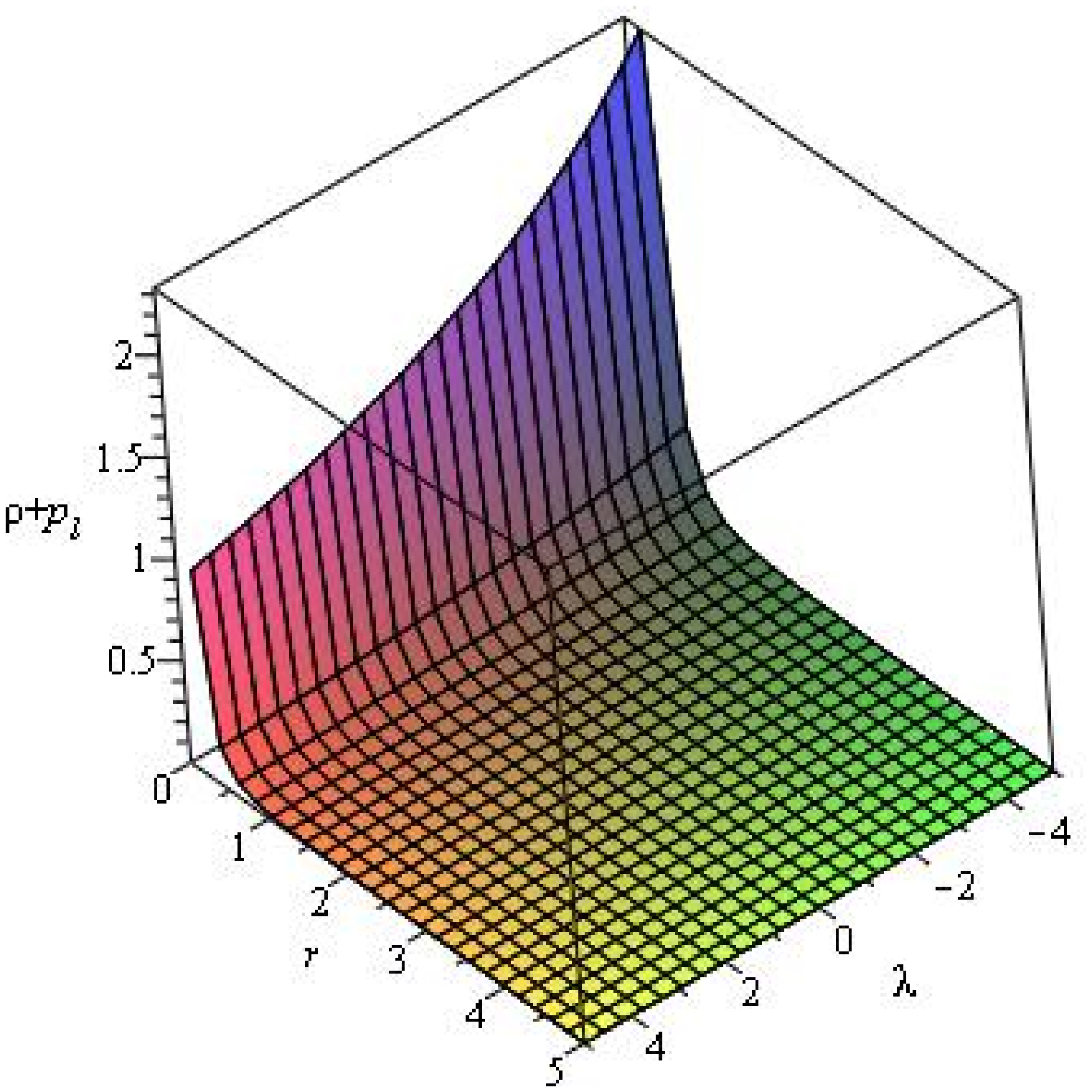}
	\caption {(a) NEC, $\rho + p_r$, (b) NEC, $\rho + p_l$.}
	
\end{figure}

\begin{figure}
	(a)\includegraphics[width=8cm, height=8cm, angle=0]{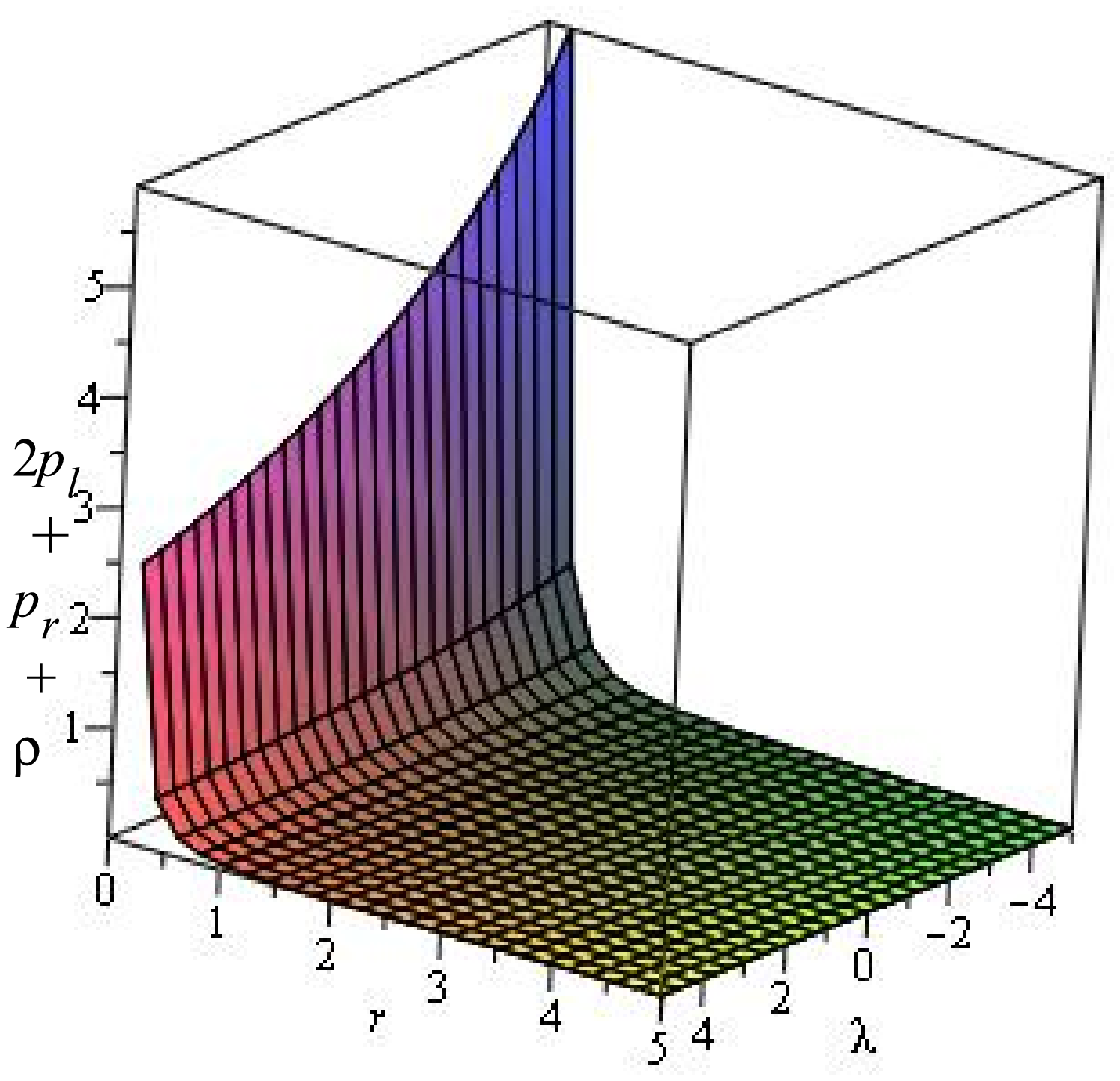}
	(b)\includegraphics[width=8cm, height=8cm, angle=0]{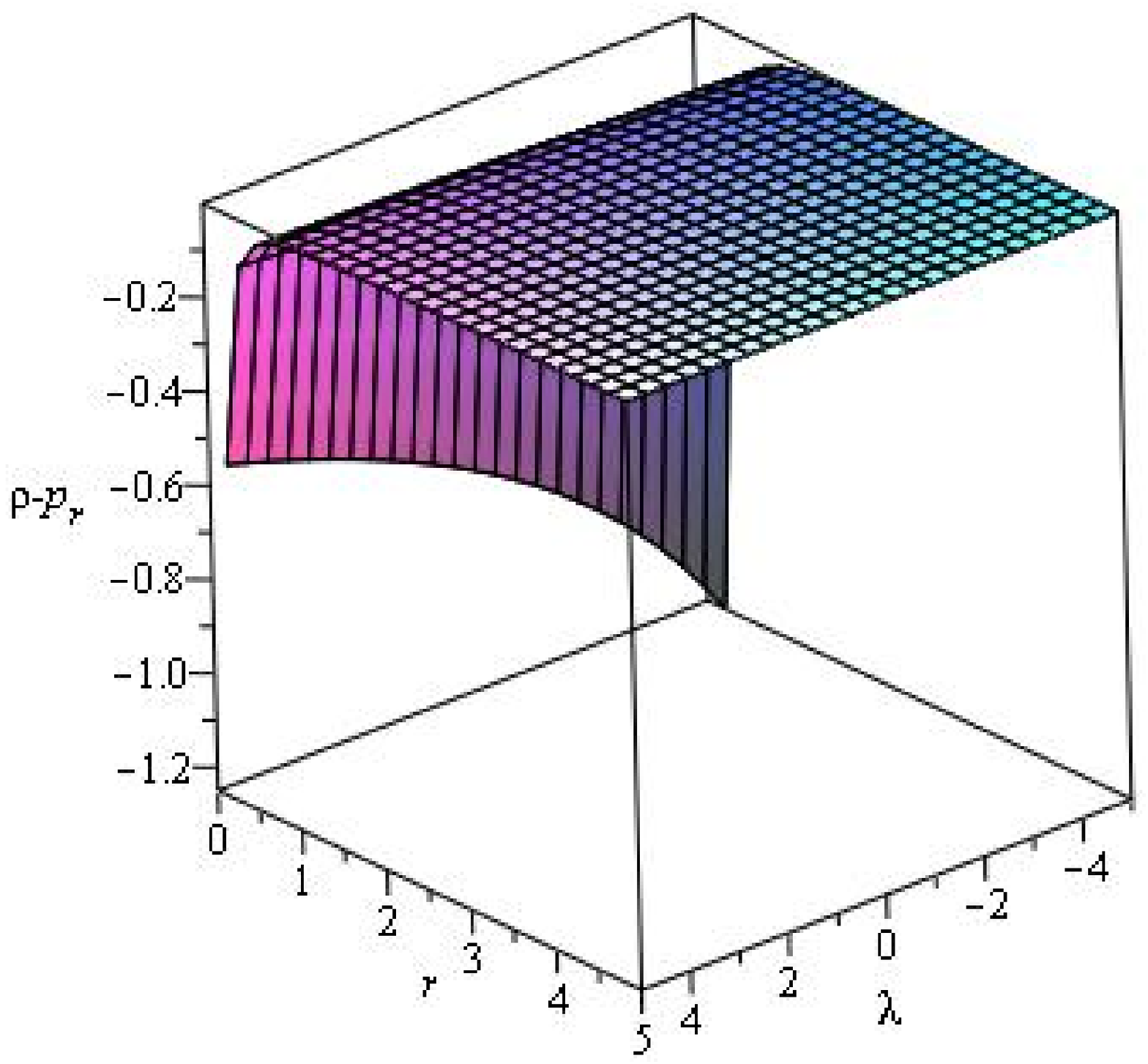}
	\caption {(a) SEC, $\rho + p_r +2p_l$ , (b) DEC, $\rho - p_r$.}
\end{figure}
\begin{figure}
	(a)\includegraphics[width=8cm, height=8cm, angle=0]{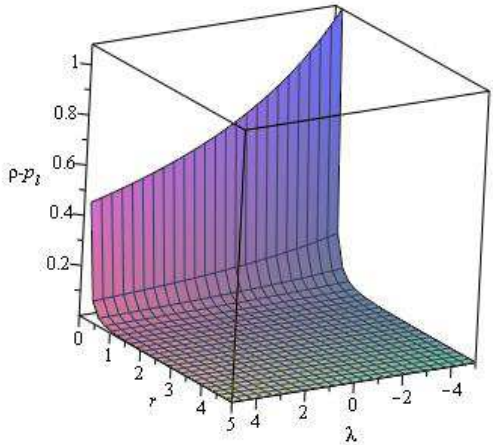}
	(b)\includegraphics[width=8cm, height=8cm, angle=0]{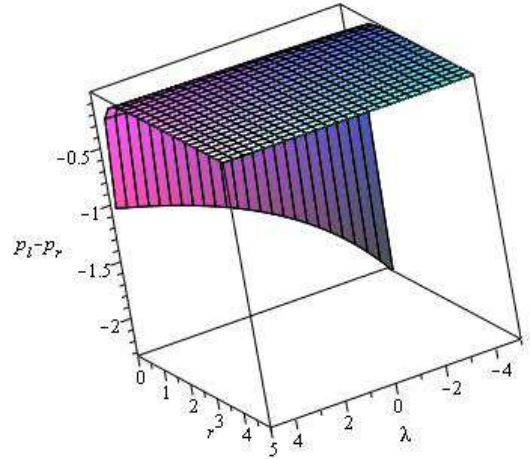}
	\caption {(a) DEC, $\rho - p_l$ , (b) DEC, $p_l - p_r$ .}
\end{figure}



\section{Discussion}

Modified gravitational theories have received considerable attention over the last few decades as a possible alternative to GR. The paramount feature of  wormholes geometries are non-vacuum solutions of Einstein's field equations. According to this theory they are filled with a matter which is different from the normal matter and is known as exotic matter. Several researchers found exotic matter to be very useful in examining whether different modified gravity models were responsible for violating energy conditions via the effective energy momentum tensors while the usual matter obey these conditions. \cite{ref44,ref55}. In the present paper we explored different models of static wormhole solutions with two different shape functions  $b(r)= \frac{r_0 \log(r+1)}{\log(r_0 +1)}$ and $b(r)= r_0 \left(\frac{r}{r_0}\right)^\gamma , \,\,\, 0<\gamma<1$ in the theory of  $f(R, T)$  gravity. One can observe from Fig. 1 and Fig. 5 that both the shape  functions satisfies throat condition, flaring out condition and typical asymptotically flatness condition for WH geometry.\\ 

Furthermore, redshift function $ U(r)$ is assumed to be constant, it indicates that a theoretical traveler's tidal gravitational force is null. In both cases we found that the geometries of the wormhole can exist even if the NEC is not violated by the usual matter i.e. it is not the exotic matter that threads the WH, but these are the extra curvature ingredients that sustained the wormhole in a modified gravity context \cite{ref55}. We analyzed the nature of SEC and WEC in the context of $f(R, T)$ gravity with two different types of shape functions to investigate this study. We emphasized the details of static wormhole geometry with variable shape function and constant redshift.\\ 

In both the models, it is observed that, for appropriate values of the parameters, we have obtain a wormhole solution with valid NEC and SEC, while  DEC  is also satisfied in terms of $p_l$ and only DEC is violated in terms of $p_r$. The presence of these solutions is in compliance with the understanding that traversable wormholes enabled by additional fundamental gravitational fields can occur without the need for exotic matter. On the other hand, the rather realistic construction needed to build a complete space-time wormhole solution may suggest that in this class of theories there are not many such solutions around.

\section*{Acknowledgments} 
The authors are thankful to GLA University, Mathura, India for providing  support and facility to carry out this research work. The authors are grateful to Ayan Banerjee, University of KwaZulu Natal, South Africa, for fruitful discussion and suggestions during his visit to GLA University.


\end{document}